\documentstyle[aps]{revtex}	  	  
\draft

\input{psfig.tex}

\begin{document}
\title{The 3-d Random Field Ising Model at zero temperature }

\author{ J.-C. Angl\`es d'Auriac \cite{byada}} 
\address{
Centre de Recherches sur les Tr\`es Basses Temp\'eratures,\\
BP 166, 38042 Grenoble, France}

\author{ Nicolas Sourlas \cite{bysourlas}}
\address{
	Laboratoire de Physique Th\'eorique 
	de l' Ecole Normale Sup\'erieure 
	\footnote  
	{Unit\'e Propre du Centre National de la Recherche Scientifique, 
	associ\'ee \`a l' Ecole Normale Sup\'erieure et \`a l'Universit\'e de 
	Paris-Sud.}
	\\
	24 rue Lhomond, \\75231 Paris CEDEX 05, France. }

\date{\today}

\maketitle

\begin{abstract}
We study numerically the zero temperature Random Field Ising Model on 
cubic lattices 
of various linear sizes $L$ in three dimensions. 
For each random field configuration we vary the ferromagnetic coupling 
strength $J$. We find that in the infinite volume limit the magnetization 
is discontinuous in $J$. The energy and its first $J$ 
derivative are continuous. 
The approch to the thermodynamic 
limit is slow, behaving like $L^{-p}$ with $p \sim .8 $ for the gaussian
 distribution of the random field. 
We also study the bimodal distribution
$h_{i} = \pm h$, and we find 
similar results for the magnetization but with 
a {\em different} value of the exponent $p \sim .6 $.
This raises the question of the 
validity of universality for the random field problem. 
\end{abstract}

\pacs{LPTENS 97/13}

\section{Introduction}
Despite its long history, the Random Field Ising Model (RFIM) is not yet 
fully undesrstood\cite{1}. It is known that in 
three dimensions and for weak 
disorder there is a paramagnetic to ferromagnetic phase transition 
but the nature of the transition  is still unknown. 
What is most remarquable is that the RFIM is 
(together with branched polymers \cite{2}) 
one of the very few cases where 
Perturbative Renormalization Group (PRG) can be
analyzed to all orders of pertubation theory\cite{3,4} 
and this analysis leads to 
wrong conclusions (while for the branched polymers the same analysis 
is correct!). 
It has been argued that this failure of PRG 
is due to replica symmetry breaking\cite{5,6,7}. This failure raises the 
more general question 
of the validity of perturbative renormalization group for disordered systems.

In the present paper we study numerically the phase transition of the 
RFIM in three dimensions. Several numerical studies have already been
performed\cite{8,9,10,11,12} but the problem has been proven to 
be very difficult and, in  
our opinion, no definite conclusion about the nature of the transition 
has been drawn yet. We simulated much  larger sizes  than before 
(up to $90^3$) with much higher statistics and we establish
new results on the phase transition of the RFIM. 
We find that the magnetization 
is discontinuous at the transition both for the gaussian and bimodal distribution. The energy and its derivative are 
continuous only for the gaussian distribution.  
The exponents are found to be non classical. Such a behaviour 
has already been  suggested on the grounds of real space renormalization group
\cite{13,14}. 
(This result has been often overlooked in the past 
 due to the approximations of the method.) We also 
find different exponents for different random field distributions, which is 
in contradiction with the PRG. 

The Hamiltonian of the RFIM is given by 

\begin{equation} 
H \  = \ -J \sum_{<i,j>} \ \sigma_{i} \sigma_{j} \ - \  \sum_{i} \ 
\ h_{i} \sigma_{i}
\end{equation}
where $ \sum_{<i,j>} $ runs over neighbouring sites of the lattice 
(we have only considered three dimensional cubic lattices with periodic 
boundary conditions) and 
$\{ h_{i} \}$ are independent random variables identically 
distributed with zero 
mean and variance one. For a given random field sample, one can vary both 
the ferromagnetic coupling $J$ and the temperature $T$, i.e. the phase 
boundary is a line in the $J$, $T$ plane. It is thought, in accordance 
with PRG,  that the nature of 
the transition and the value of the exponents do not depend on the position 
on the transition line, nor on the direction one crosses  it,
and that this is true down to zero temperature. 
So it is advantageous to work at $T=0$ where it has been shown that 
the RFIM is equivalent to the problem of finding
a maximum flow in a graph\cite{15}, for which 
very fast (polynomial) algorithms are known. 
Note that these algorithms provide the {\em exact} groud-states
and therefore there is no 
thermalization problem. Simulations using such 
algorithms have already 
been performed in the past\cite{10,11}. In the present paper we use the
latest version of 
the algorithm developped by Goldberg and Tarjan\cite{16},
which we optimized for the case of 
the cubic lattice. It has been shown \cite{16}
that this algorithm converges to the 
ground state in a time $t < L^{6} \ln L $ where L is the linear size of the 
cubic lattice. We found experimentally that $ t \sim L^{4} $\cite{17}.

\section{Gaussian distribution of random field.}
\label{s:gauss}
We first consider the case of a  gaussian distribution
of the random field with variance equal to one. It is customary 
to present the data as a function of $x=1/J$.
For every sample $\{ h_{i} \}$, we chose equally spaced $x$'s with 
$\delta x = .0125 $ and found the corresponding ground state. 
We considered lattices of 16 different linear sizes $L$, from 
$L = 7 $ to $L=90$. We studied between 750 samples for  $L=90$ 
and  20000 samples for $L=8$ and $L=7$.

\subsection{Magnetization}
\begin{figure}
\psfig{file=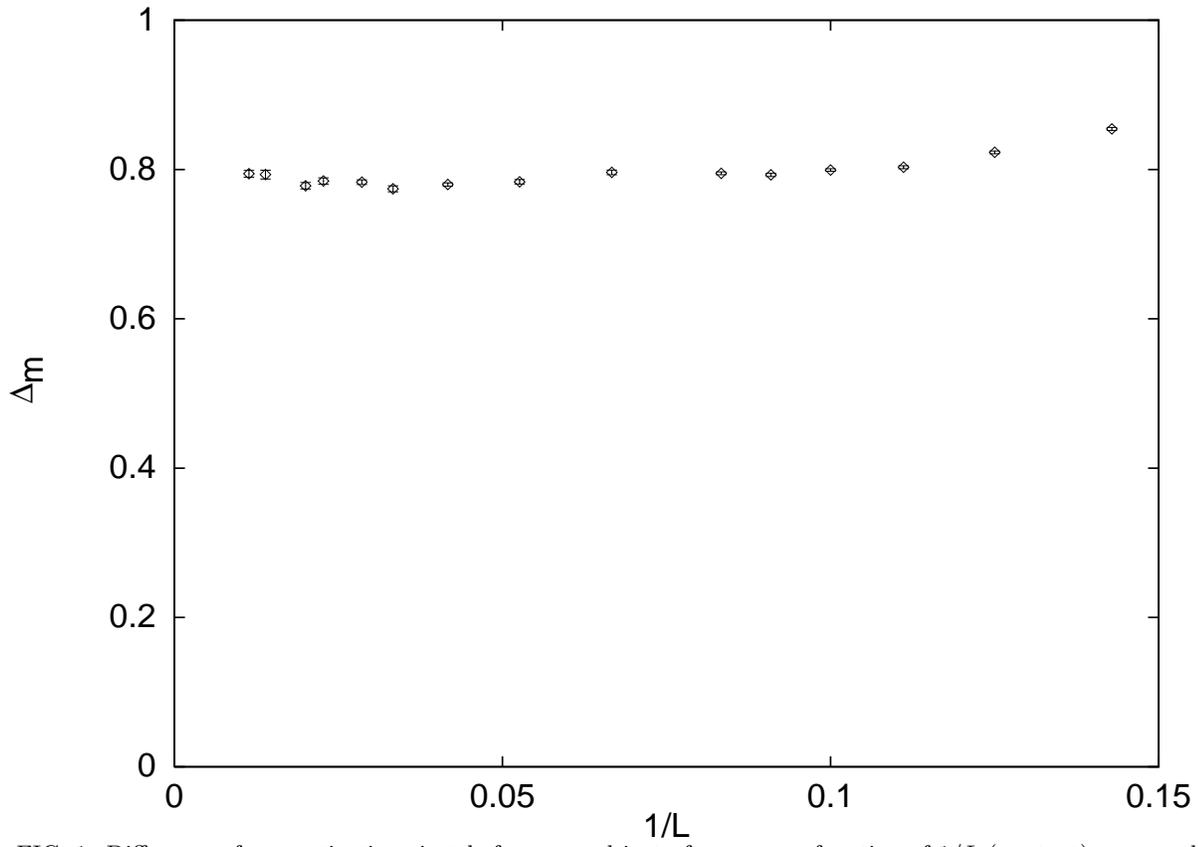,width=\hsize}
\caption{Difference of magnetizations just before $x_1$ and just 
after $x_5$ as a function of $ 1/L$ (see text), averaged over random field 
configurations drawn from a gaussian distribution. The error bars are
smaller than the symbol.}
\label{f:fig1}
\end{figure}

We study the variation of the absolute value of the magnetization $m$ 
and of the energy as a function of $x$. 
We have found that there is a region in $x$ where 
there are large discontinuities 
of $ |m| $ and that outside that region $ m $ is a smooth function of $x$. 
The amplitude of the discontinuities is volume independent (this is 
shown in Fig.~\ref{f:fig1}), while the 
width of the region in $x$ where they appear, 
shrinks as the volume increases. 
We analyze our data as follows.
For every $\{ h_{i} \}$ sample, we chose the $n_{d}$ largest variations of the 
magnetization between two succesive values of $x$, $x_{i}=1/J_{i}$ and 
$x_{i+1}$. Let's call 
$x_1 < x_2 < \cdots < x_{n_{d}} $, the values of $x$ at which they occur.
 The choice of $n_{d}$ is somehow arbitrary. We took $2 \le n_{d} \le 6 $. 
It turns out that the $x_{i}$'s fluctuate from sample to sample and their 
probability distribution is well described by a gaussian with mean 
$ \bar x_{i}(L) $ and variance $ \sigma (L)^2 $. The variance decreases with 
$L$ and is compatible with a power law decay:

\begin{equation}
\label{e:eqdelta}
\sigma (L) \sim  \sigma_{0} L^{- \delta}    
\end{equation}

\begin{figure}
\psfig{file=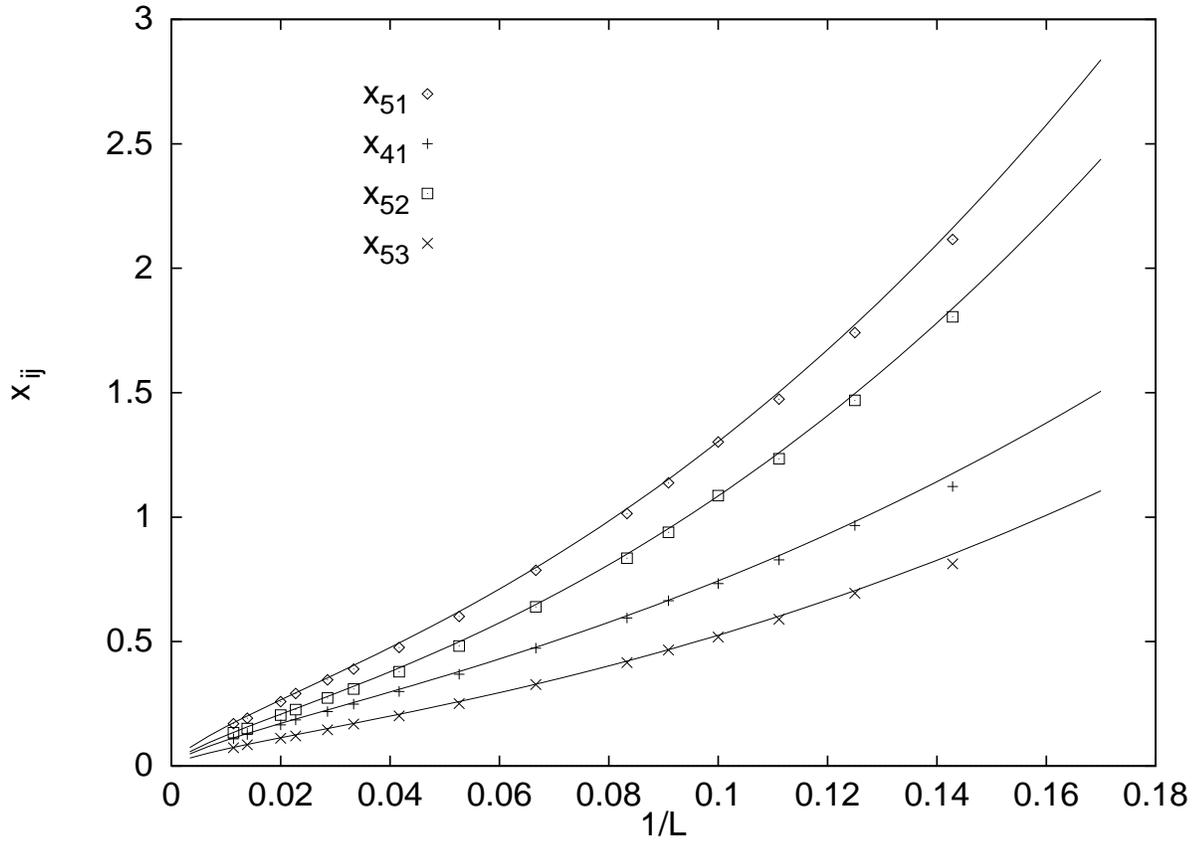,width=\hsize}
\caption{Figure 2 Averages of the position differencies $ {\bar x_{51} } $, 
$ {\bar x_{41} } $, $ {\bar x_{52}  } $, $ {\bar x_{53} } $ (see text) 
as a function of $ 1/L$ for the gaussian distribution of the random fields. 
 Continuous lines are the best fit to the data. Statistical error bars 
are smaller than the symbol size.}
\label{f:fig2}
\end{figure}

\begin{figure}
\psfig{file=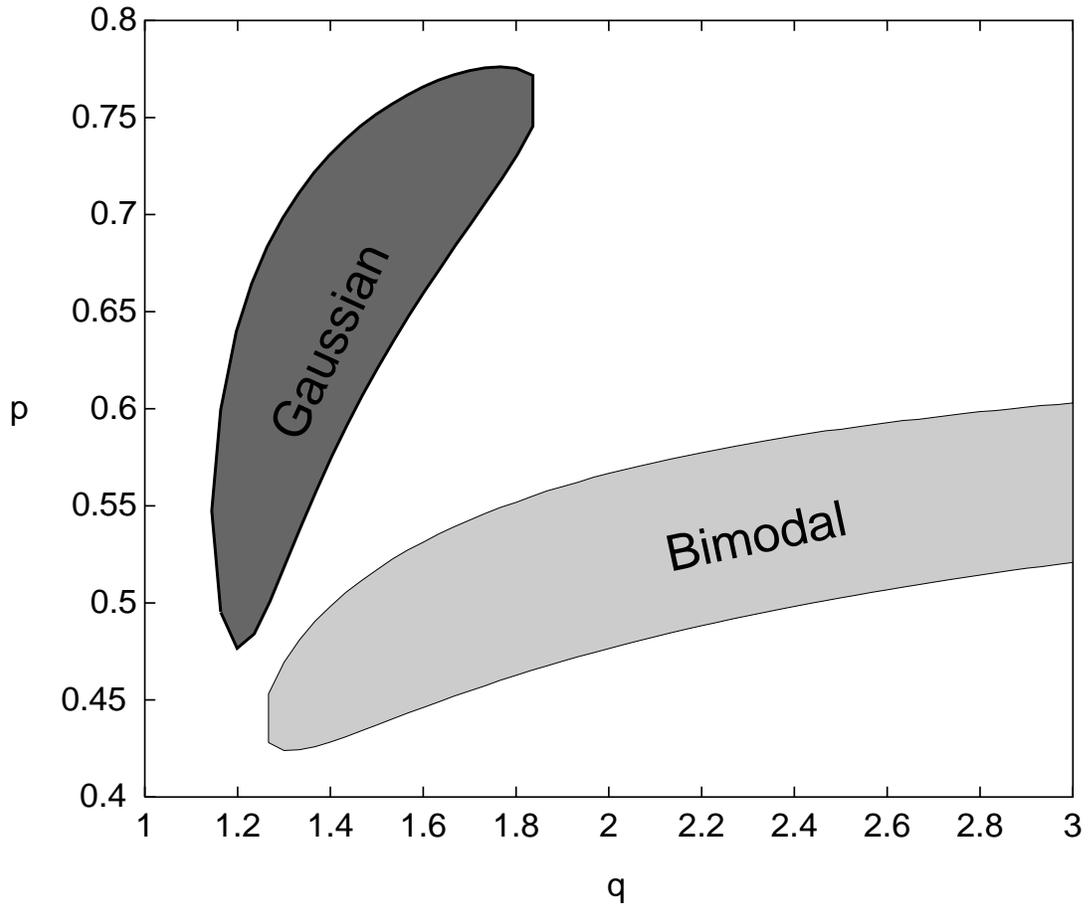,width=\hsize}
\caption{Figure 3. Allowed regions for the $p$ and $q$ exponents (see text) within  
a 90\% confidence level. The upper left region is for a gaussian distribution 
and the lower right region is for the bimodal distribution of the random 
fields.}
\label{f:fig3}
\end{figure}

We first made the ansatz that: 
\begin{equation} 
\label{e:fisimple}
\bar x_{i} \ = \ x_{\infty} \ + \ { c_{i} \over L^{p} }  
\end{equation}
i.e. that there is a single discontinuity in the infinite volume limit 
and that the approach to the thermodynamic limit obeys a power 
law. We obtain a  more accurate determination of $p$ by 
taking differences to eliminate $ x_{\infty} $ from the fit
\begin{equation} 
x_{ij} \ \equiv \ \bar x_{i}- \bar x_{j} \ = \ { c_{ij} \over L^{p} } 
\end{equation}
We have found that our results cannot be described by a single 
power for the whole range of sizes we have studied. We tried the following 
alternatives.\\
a) Analyze only the largest sizes that are compatible 
with Eq.~\ref{e:fisimple}. We found that sizes larger 
or equal to $ 24^3$ are well 
described by Eq.~\ref{e:fisimple}, with $ x_{\infty} = 2.26 \pm .01 $, 
$ p= .80 $ and a $ \chi^{2} $ per degree of 
freedom $ \chi^{2} = 1.29$\ . With 90\% probability $ .76 \le p \le .83 $ 
We also have found the value of the
exponent $\delta$  of Eq.~\ref{e:eqdelta} : 
$.78 \le \delta \le .86$ .
The choices of the lower size cut-off and of the number of discontinuities 
$n_{d} $ are quite arbitrary and induce systematic uncertainties on $p$. 
In order to get an idea on these uncertainties, we tried different 
 ``reasonable'' choices, i.e. choices with a reasonable $ \chi^{2}$ per 
degree of freedom. 
Keeping only sizes larger or equal to $ 30^3$, we found that 
$ p= .77 $ and  $ \chi^{2} = .56 $.
 With 90\% probability $ .74 \le p \le .81 $. These results 
have been obtained by analyzing the 5 largest discontinuities of 
the magnetization ($ n_{d}= 5 $). We also took
 $ n_{d}=3 $ and, in this case, $ .735 \le p \le .81 $\. 
We conclude that the 
dependence of $p$ on $ n_{d} $ is small.\\
b) Consider also a subdominant power law correction:
\begin{equation} 
x_{i} \ = \ x_{\infty} \ + \ { c_{i} \over L^{p} } 
\ (1 \ + \ { f_{i} \over L^{q} } \ )  
\end{equation}
This ansatz describes well the data down to a value 
of $L$ as small as $L=7$ with a $ \chi^{2} $
per degree of freedom $ \chi^{2} =1.$ and yields $p= .66$ and $q = 1.4$ .
There are many parameters in this fit, 
$ p $ and $q$ are correlated and the uncertainty on $p$ and $q$ 
 much larger than before. 
The best fit to the data is shown in Fig.~\ref{f:fig2}, while 
the allowed region in $p$ and $q$ is in Fig.~\ref{f:fig3}.

We conclude that the appearance of several discontinuities in the 
magnetization is a finite volume artifact and that our results are fully
 compatible with the hypothesis of a single discontinuity in the 
thermodynamic limit. The approch to this limit is a very slow power law.
The value of this power depends essentially on the assumptions underlying  
 the data analysis: 
a single power law or a power law behaviour with a subdominant 
correction. 
This discontinuity has not been seen in the previous simulations for the 
following reason. Only the average magnetization has been measured and 
for finite volumes, the position of the discontinuities 
fluctuates from sample to sample so that the average magnetization 
seems continuous. 

\subsection{Energy}
Concerning now the energy, it is convenient to separate the energy into 
two terms, $ H = - L^3 (J   H_1 + H_2) $, with 
\begin{equation}
H_1 \ = \ {1 \over L^3 } \sum_{<i,j>} \ \sigma_{i} \sigma_{j},  \quad  \  
 H_2 \ = \ {1 \over L^3 } \sum_{i} \ \ h_{i} \sigma_{i}   
\end{equation}
For a fixed sample, the value of $  H_1 = E_1 (J) $ in the ground state 
is the derivative of the energy per spin  with respect to $-J$. It can be 
shown that 
$E_1 (J) $ is a non decreasing function of $J$, i.e., $ - H $ is a 
convex function of $J$.
 We observe large discontinuities of $ H_1$ as a function of $J$, as in 
the case of the magnetization,
 but, in the present case, the amplitude of 
these discontinuities shrinks as the volume increases. 
Our data are compatible with the following ansatz:
\begin{equation} 
\label{e:delatei}
\Delta E_{i} \ \sim \ { d_{i} \over L^{a} }
\end{equation}
where $\Delta E_{i}, \ i=1, \cdots , n_{l} $ denote the $n_{l} $ largest 
discontinuities of $ E_1 $. We find $ a = .48 \pm .06 $ Therefore 
$ H_1 $, i.e. the $J$ derivative of the energy, becomes continuous at 
the thermodynamic limit.
Despite the discontinuity of 
the magnetization, the energy and its first derivative 
are continuous and the exponents are non classical. 
Eq.~\ref{e:delatei} 
sheds some light to the long standing question of the 
existence of different spin configurations almost degenerate in energy.
The reasonning goes as follows.  
For $ J = 0 $, $ E_1 \sim .0 $, while for $ J \gg 1. $, $ E_1 = 3. $ 
In general $ E_1 $ is a slow varying monotonous function of $ J $ except 
in the region $ .2 < J < .5 $ where most of the variation occurs. 
If the spin configuration $ \sigma_{i}^{(1)} $ is a ground state for 
$ J = J^{(1)} $ with energy per spin $- E^{(1)} = J^{(1)} E_{1}^{(1)} +
 E_{2}^{(1)} $ and another 
 spin configuration $ \sigma_{i}^{(2)} $ is a ground state for 
$ J = J^{(2)} > J^{(1)} $ with energy per spin $ - E^{(2)} = J^{(2)} E_{1}^{(2)}
 + E_{2}^{(2)} $ then for $ J^{(3)} = (E_{2}^{(2)} - E_{2}^{(1)} )
/ (E_{1}^{ (1) } - E_{1}^{ (2) } ) $, the two configurations have exactly 
the same energy $ - E^{(3)} = J^{(3)} E_{1}^{(1)} +
 E_{2}^{(1)} $ ($J^{(1)} < J^{(3)} <J^{(2)} $ because of the convexity 
of $ - H $ )\footnote{One could find the ground state energy 
for $J=J^{(3)}$, which is either identical to 
$ E^{(3)} $ or lower. In the former case the ground state energy is exactly 
known in the entire range $ J^{(1)} \le J  \le J^{(2)} $. In the latter 
case one could iterate this procedure and so it is possible to compute  
the ground state energy exactly as a function of $J$. 
This will be exploited in a forthcoming publication.}.
If the largest discontinuity is of the order of $ L^{- \alpha } $, 
it follows that in the interval 
$ J^{(1)} < J < J^{(2)}$ there are  at least 
$ \sim ( E_{1}^{(2)} - E_{1}^{(2)} ) L^{a} $ different 
configurations  with energy $ E_2  \le  E \le E_1 $, 
which are pairwise exactly degenerate for some value of $ J $.
This argument is valid for any choice of $ E_1 $, $ E_2 $, 
arbitrarly close to each other.

\subsection{Disconnected susceptibility}
In order to compare with previous simulations, we have considered the 
so called disconnected susceptibility $ \chi_{dis} = L^{3} \overline{ m^{2}} $ 
i.e. the square of the magnetization averaged over the disorder.
It has been assumed that at zero temperature 
\begin{equation} 
\label{e:kidis}
\chi_{dis} \sim L^{4 - \bar \eta} f ( (x- x_{c}) L^{ 1 / \nu} ) 
\end{equation}
where $ x_{c} $ is the location of the transition at infinite volume and 
$ \nu $ the correlation length exponent. In \cite{11} it was found that, 
for gaussian random fields at zero temperature, 
$\bar \eta =1.1 \pm .1 $ and $\nu =1 \pm .1$.
In fact Eq.~\ref{e:kidis} is compatible with our results.
To see this, we propose the following  
simplified model:  $m =m_{1} $ for 
$ x < x_{1} $ and $m =m_{2} $ for $ x > x_{1} $, where  
$ x_{1} $ is a gaussian random variable with mean 
$ \bar x  \sim x_{\infty} +  c / L^{p}$ and 
standart deviation $ \sigma \sim \sigma_{0} L^{- \delta}  $ . 
A simple calculation gives, in this model, 
$ \chi_{dis} \sim  L^{3}  f ( (x- x_{\infty}) L^{  \delta} +
L^{ \delta - p } ) $. If $ \delta = p $,  
this agrees with Eq.~\ref{e:kidis}  
provided $ \bar \eta = 1 $ (which is compatible with \cite{11}) 
and $ \delta = 1/ \nu $.  
We found above (case a), that $ \delta = .82 \pm .04 $ 
and $ p = .80 \pm .04 $, 
compatible with $ \delta = p $. This gives $\nu = 1.22 \pm .06 $ which is 
slightly different from the value $\nu =1 \pm .1$  of \cite{11} .  
If we restrict our analysis to the same 
(smaller) sizes as in \cite{11}  we find $ \delta = .96 \pm .03 $, i.e. $\nu =1.04 
\pm .03 $, as in \cite{11} 
and $ p = 1.17 \pm .03 $. Alternatively we also have fitted 
directly our data using Eq.~\ref{e:kidis}. 
This provides a different method of analysis. 
Since, as already mentionned, a single power is 
not sufficient to describe the whole range of sizes, we
have restricted our analysis to the largest sizes.
We have found $ \bar \eta = .98 \pm .01 $ in agreement with our simple model 
and $ x_{\infty} =2.265 \pm .005 $, a value compatible with the
previously deduced value.

\section{Bimodel distribution of random field.}
\label{s:bimod}
We consider next the case of a bimodal field distribution, 
\begin{equation}
P(h_i) = (1/2) \left( \delta(h_i-1) + \delta(h_i+1) \right)
\end{equation}
i.e. 
$ h_{i} = \pm 1 $ with equal probability. 
For every sample $\{ h_{i} \}$, we chose equally spaced $x$'s with 
$\delta x = .0125 $ and we scanned in $x$.
 The behaviour of the 
magnetization is very similar to the case of the gaussian probability 
distribution of the random field. 
We found again that there is a region in $x = h/J $ where there are 
large discontinuities 
of $ |m| $.
The amplitude of the discontinuities is volume independent, while the 
width of the region in $x$ where they appear, shrinks as the volume increases.
We made the same hypothesis as in the case of the gaussian distribution,
namely \\
a) We found that sizes larger or equal to $ 24^3$ are well 
described be equ. (3), with $ x_{\infty} = 2.21 \pm .01 $, 
$ p= .60 $ and a $ \chi^{2} $ per degree of 
freedom $ \chi^{2} = .6 $. With 90\% probability $ .57 \le p \le .63 $. 
Keeping only sizes larger or equal to $ 30^3$, we found that 
$ p= .59 $ and  $ \chi^{2} = .76 $.
 With 90\% probability $ .55 \le p \le .63 $ These results were obtained by 
analyzing the 4 largest discontinuities of the magnetization ($ n_{d}= 4 $). \\
b) Considering also a subdominant power correction
the data are well described  with a $ \chi^{2} $
 per degree of freedom $ \chi^{2} =1.2  $ and yields $ p= .56 $ and $ q = 3. $ 
Unfortunatly the error bars are much larger than before. 
The allowed region in $p$ and $q$ is presented in Fig.~\ref{f:fig3}.

For the energy, the situation is more complex. In addition to the type of 
singularities we observed in the gaussian case, there are new stronger 
singulaties, at integer values of $x$, i.e. $x=3,4,5,...$ These singularities 
are present for all sizes and the amplitude of the discontinuity 
of $E_1$ at these points seems to be volume independent. Nothing in 
particular is happening to the magnetization at these points.
While the presence of these singularities 
is intuitively easy to understand 
in the case of the bimodal distribution, 
the decoupling of the magnetization 
from the energy is less intuitive. 

\section{Conclusion and Discussion}
In this paper we have found that he exponent $p$,  
which describes the finite size corrections and is 
usually identified with $1/ \nu $ ( $\nu $ is the 
correlation length exponent), seems to take different values for
 the gaussian or 
the bimodal distribution of the random field. This would signal the 
breaking of universality for the RFIM. 
It is therefore important to discuss 
 a) the uncertainties on the determination of $p$ and 
 b) the identification of $p$ with $1/ \nu$.\\
a) The statistical errors on $p$ are rather small. 
The largest uncertainties 
are due to the assumptions made to analyze the data, namely 
single power law behaviour or inclusion of subdominant corrections in $L$.  
But whatever the assumptions are, one finds significally different values, 
provided one uses the same assumptions for both random field 
distributions. \\
b) Let us remind the arguments which lead to the identification of $p$ 
with $1/ \nu $ 
For finite systems of linear sizes $L$, one can define an
 effective critical temperature $ T_{eff} (L)$. 
For a second order transition, $ T_{eff} (L)$ could be the temperature
 for which the susceptibility  is maximum. 
The finite size scaling hypothesis  
 asserts that  $ T_{eff} (L)$  is shifted 
 from $T_{c}$, the critical temperature of the infinite volume 
system: $ T_{eff} = T_{c} + c L^{- 1/ \nu} $. 
Assume a second order transition 
 for the RFIM and, in agreement with 
PRG, that the exponents do not depend 
on the position on the transition line 
in the $T$ $J$ plane, or the direction one crosses it. Assume  
furthermore, that this is valid down to zero temperature. 
We then expect the locations of the magnetization discontinuities 
$x_{i} (L) $'s to be shifted 
from their infinite volume value $x_{\infty } $: $x_{i} (L) = 
x_{\infty } + c_{i} L^{- 1 / \nu} $, i.e. 
that the exponent $p$ defined above 
is $ p = 1 / \nu $ and therefore is universal. 
If $p$ is different for the gaussian or the bimodal 
random field distribution as we claimed above, one of the above 
generally accepted assumptions, which can be derived 
in the context of PRG, in not valid 
for the random field Ising model. 
After all it is already 
known that the $\epsilon$ expansion leads to 
incorrect results for the RFIM.
The most conservative attitude would be to assume (without any a priori 
reason) that the bimodal distribution is singular at $T=0$ 
and that the general argument does not apply.
But we would like to point out that, except from aesthetic appealing, 
the only theoretical argument for universality in disordered 
systems is based on (unbroken) replicas and perturbative renormalization 
group. And we know that this may be incorrect. We would like to point out 
that also in the spin-glass case, there is numerical evidence that 
universality may be broken\cite{18}). 
We feel therefore that universality should be 
further checked for disordered systems. 
In particular it would be interesting 
to study other distributions of random fields.

\end{document}